\documentclass[useAMS,usenatbib]{mn2e}
\pdfoutput=0

\usepackage{graphicx}
\usepackage{aasmacros}
\usepackage{color}
\usepackage{epsfig}
\usepackage{enumitem}

\def\hmpc{$h^{-1}$Mpc}

\def\msol{M$_\odot$}
\def\hmsol{$h^{-1}$M$_\odot$}

\def\om{\Omega_m}

\def\omb{\Omega_b}
\def\s8{\sigma_8}

\def\lcdm{$\Lambda$CDM}

\def\x2{$\chi^2$}
\def\hmsol{$h^{-1}\,$M$_\odot$}

\def\NNm1{\langle N(N-1) \rangle}

\def\m_star{M_\ast}

\def\rhob{\tilde{\rho}}

\def\lcdm{$\Lambda$CDM}

\def\om{\Omega_m}

\def\omb{\Omega_b}
\def\s8{\sigma_8}

\def\hmpc{$h^{-1}\,$Mpc}

\def\x2{$\chi^2$}
\def\hmsol{$h^{-1}\,$M$_\odot$}

\def\NNm1{\langle N(N-1) \rangle}

\def\p0{P_0(r)}

\def\fq{f_{\rm Q}}

\def\mgal{M_\ast}

\def\dpx{\delta_{\rm gal}}
\def\mdot{\dot{M}_h}
\def\mhalo{M_{h}}

\def\zhalf{z_{1/2}}
\def\psat{P_{\rm sat}}
\def\rexp{R_{\rm exp}}
\def\ns{n_{S}}
\def\rhop{\delta_{\rm gal}}
\def\ssfr{\rm sSFR}
\def\msol{{\rm M}_\odot}
\def\fmerge{f_{\rm merge}}
\def\cvir{c_{\rm vir}}
\def\rhob{\delta_{\rm gal}}

\title[Halo and Galaxy Histories]{Halo Histories vs. Galaxy Properties at z=0, III: \\The Properties of Star-Forming Galaxies}
\author[Tinker et.~al.]{\parbox{\textwidth}{Jeremy L. Tinker$^1$, ChangHoon Hahn$^1$,   Yao-Yuan
  Mao$^{2}$,  Andrew R. Wetzel$^{3,4,5}$\thanks{Caltech-Carnegie Fellow}}\\
\footnotesize
  $^1$Center for Cosmology and Particle Physics, Department of
  Physics, New York University, New York, NY, USA\\
$^2$Department of Physics and Astronomy \& Pittsburgh Particle Physics, Astrophysics, and Cosmology Center (PITT PACC)\\
$^3$The Observatories of the Carnegie Institution for Science, Pasadena, CA, USA \\
$^4$TAPIR, California Institute of Technology, Pasadena, CA, USA \\
$^3$Department of Physics, University of California, Davis, CA, USA }

\begin{document}


\pagerange{\pageref{firstpage}--\pageref{lastpage}} \pubyear{2016}

\maketitle

\label{firstpage}

\begin{abstract}

  We measure how the properties of star-forming central galaxies
  correlate with large-scale environment, $\delta$, measured on $10$
  \hmpc\ scales. We use galaxy group catalogs to isolate a robust
  sample of central galaxies with high purity and completeness. The
  galaxy properties we investigate are star formation rate (SFR),
  exponential disk scale length $\rexp$, and Sersic index of the
  galaxy light profile, $\ns$. We find that, at all stellar masses,
  there is an inverse correlation between SFR and $\delta$, meaning
  that above-average star forming centrals live in underdense
  regions. For $\ns$ and $\rexp$, there is no correlation with
  $\delta$ at $\mgal\la 10^{10.5}$ $\msol$, but at higher masses there
  are positive correlations; a weak correlation with $\rexp$ and a
  strong correlation with $\ns$. These data are evidence of assembly
  bias within the star-forming population. The results for SFR are
  consistent with a model in which SFR correlates with present-day
  halo accretion rate, $\mdot$. In this model, galaxies are assigned
  to halos using the abundance matching ansatz, which maps galaxy
  stellar mass onto halo mass. At fixed halo mass, SFR is then
  assigned to galaxies using the same approach, but $\mdot$ is used to
  map onto SFR. The best-fit model requires some scatter in the
  $\mdot$-SFR relation. The $\rexp$ and $\ns$ measurements are
  consistent with a model in which both of these quantities are
  correlated with the spin parameter of the halo, $\lambda$. Halo spin
  does not correlate with $\delta$ at low halo masses, but for higher
  mass halos, high-spin halos live in higher density environments at
  fixed $\mhalo$. Put together with the earlier installments of this
  series, these data demonstrate that quenching processes have limited
  correlation with halo formation history, but the growth of active
  galaxies, as well as other detailed galaxies properties, are
  influenced by the details of halo assembly.

\end{abstract}


\section{Introduction}

This is the third installment of a series of papers focused on
possible connections between the properties of present-day galaxies
and the evolutionary histories of the halos in which those galaxies
formed. In each work, we select a sample of `central' galaxies with
which to make our comparisons. These galaxies live at the center of
distinct halos---these galaxies could also be referred to as `field
galaxies'---and have not been subjected to the type of physical
processes that transform galaxies that have been accreted as
satellites onto groups and clusters. In Papers I and II, we
investigated the quenched fraction of central galaxies in the SDSS,
$\fq$, comparing various measurements of this quantity to models in
which halo formation history is correlated with mean stellar age,
known as the age-matching model (\citealt{hearin_watson:13,
  hearin_etal:15}). This model predicts that red-and-dead galaxies
live in the oldest halos, while the most active star-formers live in
the youngest. In Paper I, we determined that such models predict a
dependence of $\fq$ on large-scale density that is inconsistent with
observations.  At fixed mass, halo clustering depends on halo age, an
effect known as assembly bias (\citealt{wechsler_etal:06,
  gao_white:07}). Thus, the age-matching model predicts that quiescent
central galaxies should predominantly live in dense regions, but will
rarely be found in underdense regions. The measurements of Paper I
shows essentially no correlation between $\fq$ and density at the halo
masses where the age-matching model predicts it to be the strongest.

In Paper II we explored $\fq$ through galactic conformity (see, e.g.,
\citealt{kauffmann_etal:13, hearin_etal:15}), once again finding that
halo formation has a limited, if any, role in determining whether a
galaxy makes the transition from star-forming to quiescence. In this
paper, we narrow our sample to only looking at central galaxies that
are actively star-forming. Theoretical models of galaxy growth inside
halos usually assume some relationship between the properties of
disky, active galaxies, and the dark matter halo that surrounds
them. We will test several of these assumptions.

In some respects, the correlation between galaxy growth and halo
growth is undeniable: larger galaxies live in larger halos. The
abundance matching model has been used as a function of redshift to
infer the details of this correlation (\citealt{conroy_wechsler:09,
  behroozi_etal:13, behroozi_etal:13_letter, moster_etal:13}). In all of
these models, it is assumed that the baryonic accretion rate onto a halo is
proportional to the dark matter accretion  rate of that
halo. Abundance matching tells one how much the galaxy within a given
halo has grown, and with this information one can infer the efficiency
of star formation over that time interval. For two halos of the same
mass today, the halo that grew the most over that time should have the highest
star formation rate. Halo growth rate is tied to closely tied to
assembly bias, thus this prediction of the abundance matching ansatz
creates testable predictions for the population of central galaxies.

In the canonical picture of galaxy formation in a CDM universe, the
properties of disk galaxies are determined by the relationship between
dark matter and baryons. Accreted baryons are converted into a disk of
cold gas and stars that has an exponential scale length determined by
the angular momentum of the dark matter halo, which is aligned and
distributed proportionately with the baryonic material
(\citealt{dalcanton_etal:97, mo_etal:98}). Recent cosmological
hydrodynamic simulations have found that halo spin is correlated
with the size and morphology of the stellar material, but with significant
scatter (\citealt{teklu_etal:15, zavala_etal:16,
  rodriguez_gomez_etal:17}). Several studies have shown
that halo angular momentum (or `spin' for brevity) is another halo
property that influences---or is influenced by---the clustering of the
halos: higher spin halos live in more dense regions
(\citealt{gao_white:07, bett_etal:07}), giving us the opportunity to
test this theory observationally. 

Cosmological hydrodynamic simulations demonstrate that merger activity
is strongly correlated with the buildup of a central bulge (see, e.g.,
\citealt{brooks_christensen:16} and citations therein). The merger
rate of dark matter halos depends on large scale density such that
more mergers occur in higher density environments. This dependence is
not particularly strong---the merger rate increases by a factor of
$\sim 2$ over roughly a factor of 10 in $\rho$
(\citealt{fakhouri_ma:09}). But bulgeless, disk-dominated galaxies in
the local universe have most likely experienced the lowest amount of
merging in the galaxy population. If so, they should reside in the
lowest densities within which such galaxies can be found, making it
possible to detect this effect.

The key to all of these supposedly observable trends, as always, is
having an unbiased sample of central galaxies. Galaxies that orbit
within larger halos as satellites have been subject to a set
of physical processes that are distinct from those than can act on
central galaxies in the field. As in Papers I and II, we will use
group catalogs to identify central galaxies within the full SDSS
DR7. The tight correlation between stellar mass and halo mass implies
that we can use stellar mass as a reasonable proxy for halo
mass. Thus, a pure and complete set of central galaxies at fixed
stellar mass is an effective way to examine a set of halos at fixed
dark matter mass. In this context, the search for assembly bias is
much cleaner and more straightforward.

Throughout, we define a galaxy group as any set of galaxies that
occupy a common dark matter halo, and we define a halo as having a
mean interior density 200 times the background matter density. For
all distance calculations and group catalogs we assume a flat, \lcdm\
cosmology of $(\om,\s8,\omb,n_s,h_0)=(0.27, 0.82,
0.045,0.95,0.7)$. Stellar masses are in units of $\msol$.

\begin{figure*}
\psfig{file=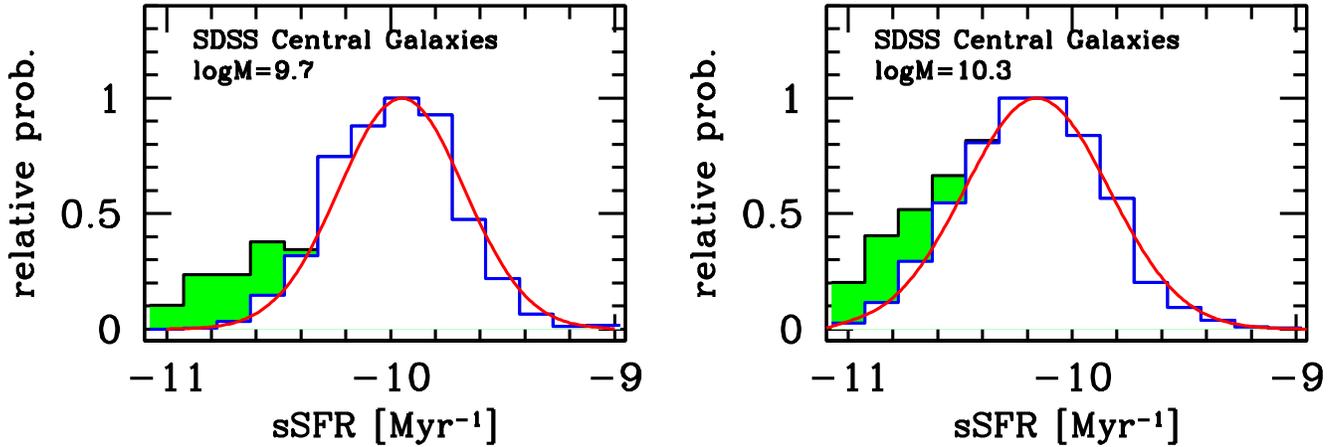,width=1\linewidth}
\vspace{-11cm}
\caption{ \label{mdot_histo} {\it Left Panel:} The distribution of
  specific star formation rates (sSFR) for central galaxies with
  $\log\mgal=9.7$. The red curve is a lognormal fit to the data at the
  mode and right-hand side of the distribution. We assume that the
  star-forming main sequence is a symmetric distribution about the
  mean, implying that the galaxies in excess of the lognormal at low
  sSFR values are `transitional' galaxies in the green valley. {\it
    Right Panel:} Same as the left, but now for galaxies with
  $\log\mgal=10.3$}
\end{figure*}

\section{Data, Measurements, and Methods}

\begin{figure*}
\psfig{file=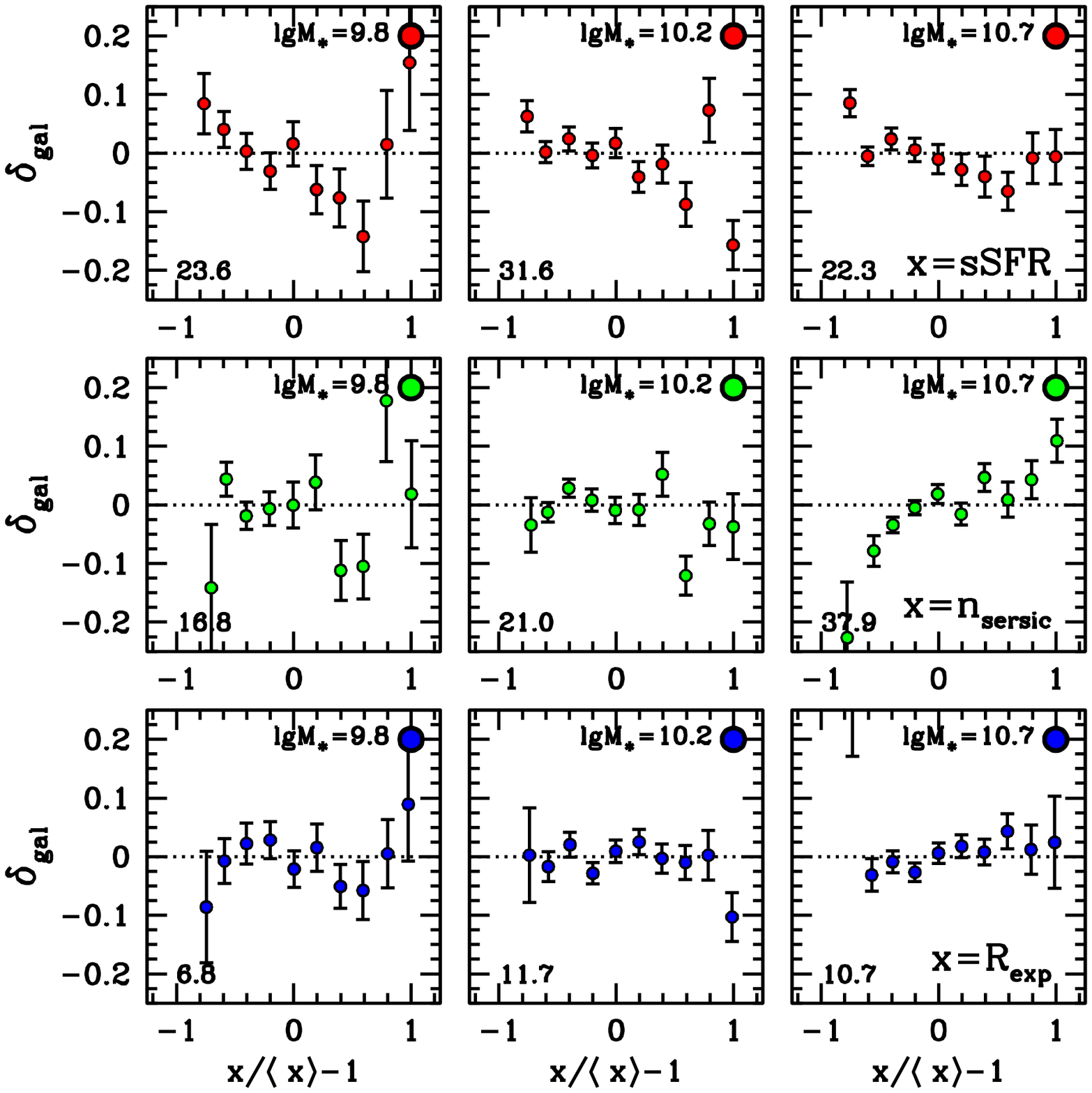,width=1\linewidth}
\caption{ \label{mdot_inv_all} The correlation between large-scale
  density, $\rhop$, and properties of star-forming galaxies. From
  bottom to top, the galaxy properties in each row are $\rexp$, the
  scale length of the exponential fit to the galaxy light profile,
  $\ns$, the best-fit sersic index of the galaxy light profile, and
  $sSFR$, the specific star formation rate. The columns represent
  different galaxy stellar masses, from low to high, as indicated in
  the panels.  }
\end{figure*}

\subsection{NYU Value-Added Galaxy Catalog and Group Catalog}

As in Papers I and II, we use the NYU Value-Added Galaxy Catalog
(VAGC; \citealt{blanton_etal:05_vagc}) based on the spectroscopic
sample in Data Release 7 (DR7) of the Sloan Digital Sky Survey (SDSS;
\citealt{dr7}).  We use stellar masses from the {\tt kcorrect} code of
\cite{blanton_roweis:07}, which assumes a \citet{chabrier:03} initial
mass function. Estimates of the specific star formation rates (sSFR)
of the VAGC galaxies are taken from the MPA-JHU spectral
reductions\footnote{\tt http://www.mpa-garching.mpg.de/SDSS/DR7/}
(\citealt{brinchmann_etal:04}).

The group catalogs are created from volume-limited stellar mass
samples. Details of the group finding process can be found in
\cite{tinker_etal:11} and further tested in
\cite{campbell_etal:15}. In brief, the group finding algorithm used
here is based on that of \cite{yang_etal:05}, in which the full galaxy
population can be decomposed into two distinct populations: central
galaxies, that exist at the center of a distinct dark matter halo, and
satellite galaxies, that orbit within a larger dark matter halo. Each
galaxy in the sample is given a probability of being a satellite
galaxy, $\psat$. In our fiducial sample, galaxies with $\psat\ge 0.5$
are classified as satellites, while galaxies with $\psat<0.5$ are
classified as centrals.

In this paper we focus exclusively on central galaxies. Impurities and
incompleteness are inevitable consequences of any group-finding
process. Using our group finer, the purity of the full sample of
central galaxies is around 90\%, with a completeness of $\sim
95\%$. However, the purity of the sample of central galaxies has a
strong correlation with $\psat$. The vast majority of central galaxies
have $\psat<0.01$, with many being exactly 0. Most incorrectly
classified central galaxies---i.e., true satellite galaxies that are
labeled as centrals by the algorithm---have $\psat$ in the range
$0.01<\psat<0.5$. Thus, we can create a `pure' sample of central
galaxies by reducing the $\psat$ threshold to $\psat<0.01$. This
excludes roughly $\sim 15\%$ of classified centrals but reduces the
impurity to $\sim 1\%$. Our fiducial results in this paper will use
samples of pure central galaxies in order to avoid any bias from
including true satellite galaxies in the sample. In Appendix B we
demonstrate that our fiducial results are largely unaffected by this
choice.

In addition to focusing on central galaxies, we specifically want to
investigate the properties of galaxies on the star-forming main
sequence (SFMS). The SFMS is characterized by a power-law dependence
of star-formation rate (SFR) and $\mgal$, with a lognormal scatter of
roughly 0.3 dex around the mean $\log$SFR
(\citealt{noeske_etal:07a}). Dividing a sample of galaxies into star
forming and quiescent usually involves splitting a bimodal
distribution at the minimum between the two modes of the galaxy
distribution. However, there are galaxies that are on either side of
that division that are not canonical star forming or quiescent
objects, but rather in the process of migrating from the former to the
latter. We call these transitioning galaxies. The relative height of
this `green valley' to the peaks contains information about the
quenching timescale of galaxies. Using this information,
\cite{wetzel_etal:13_groups2} and \cite{hahn_etal:17_tq} find that satellite
galaxies and central galaxies typically spend $\sim 2$ and $\sim 4$
Gyr in this migration, respectively. 

Identifying which galaxies are transitioning and which are merely
below-average star-formers is not possible with the dataset we use
here. We thus require a procedure to statistically account for the
fact that some fraction of the population is not on the canonical
SFMS. Figure \ref{mdot_histo} shows the distribution of specific star
formation rate (sSFR$\equiv {\rm SFR}/\mgal$) for low and high-mass
central galaxies. The red curves show a lognormal fit to the
distribution, but only using the data rightward of the mode of the
distribution. We make the assumption that the true SFMS is a symmetric
lognormal distribution. The area of the histogram above the red curve
shows the fraction of galaxies that are assumed to be
transitioning. Galaxies in this range of sSFR are weighted by the
ratio of the red curve to the total histogram. This procedure has two
benefits: (1) the sample of galaxies has a true lognormal distribution
of sSFR, thus making it straightforward to create theoretical models
that connect halo accretion rate to galaxy SFR (which we will discuss
in \S \ref{s.sims}. (2) If transitioning galaxies occupy any special
environment, this will not impact our results. However, we show in
Appendix B that, in fact, using all galaxies does not change our
results.

In addition to SFR, we utilize two other properties of central
galaxies in SDSS; their exponential scale lengths, $\rexp$, and the
Sersic indices of their light profiles, $\ns$. We obtain the values of
$\rexp$ from the NYU-VAGC. For $\ns$, we use updated values from the
NASA-Sloan Atlas (NSA\footnote{The NSA is made publicly available by
  M.~R.~Blanton at {\tt http://www.nsatlas.org}. We use here the
  version of the NSA updated for target selection of the MANGA Survey
  (\citealt{manga}), which extends to upper redshift limit to
  $z=0.15$, which includes all galaxies in our group catalogs.} The
value of $\rexp$ is the value of the exponential scale length in a
pure exponential model fit to the galaxy magnitude profile. The value
of $\ns$ is determine by fitting the magnitude profile to a Sersic
function with the form $I(r)=A\exp\left[ -(r/r_0)^{1/\ns}\right]$. For
a purely exponential disk, $\ns=1$, while for a purely de Vaucouleurs
profile $\ns=4$. In \cite{blanton_etal:05}, galaxies with blue $g-r$
colors exhibit $\ns$ values in the range $0.5$-$2.5$. Since our sample
of central galaxies all lie on the SFMS, the vast majority will have
some disk component. The use of $\ns$ is a proxy for how
bulge-dominated the galaxy is.

\begin{figure*}
\psfig{file=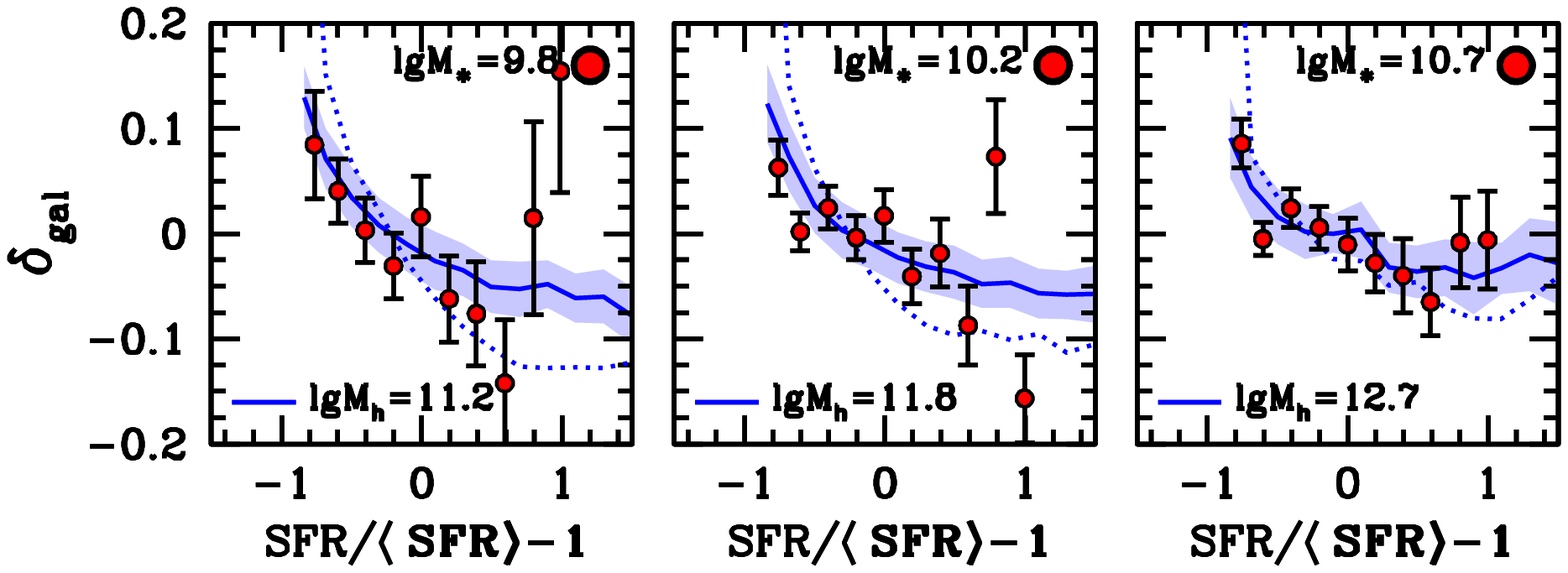,width=1\linewidth}
\vspace{-11cm}
\caption{ \label{mdot_sfr_inv} Comparison between measurements and
  models for the correlation between sSFR and $\rhop$ for central
  galaxies.  The points with errorbars are the same as those presented
  in Figure \ref{mdot_inv_all}. The curves are the results from halo
  abundance matching models. The dotted line is a model in which there
  is no scatter between $\mdot$ and sSFR. This model yields results
  that agree with the general trends of the data, but the slope of the
  correlation between sSFR and $\rhop$ is notably steeper. The solid
  curve is a model which incorporates 0.25 dex of scatter in
  $\log$~sSFR at fixed $\mdot$. This value of scatter yields the best
  $\chi^2$. The shaded region is the uncertainty in the model through
  jackknife sampling of the simulation volume.}
\end{figure*}

\subsection{Measuring Large-scale Environment}
\label{s.galden}

For each galaxy, we estimate the large-scale environment by counting
the number of neighboring galaxies within a sphere of radius 10 \hmpc\
centered on each galaxy. This is the same definition of environment
used in Paper I. In each volume-limited sample, we use the full galaxy
sample down to the absolute $r$-band magnitude limit to calculate
galaxy density. We use this sample, rather than the stellar mass
volume-limited sample, simply because the magnitude-limited sample has
more galaxies and thus reduces shot noise in the measurement. The
effect of peculiar velocities is small, and the 10 \hmpc\ scale is a
clear distinction from the density on the scale of the halo virial
radius. We use the {\tt mangle} softward of \cite{swanson_etal:08} to
characterize the SDSS survey geometry and create random samples.

Rather than use the absolute number of galaxies around each object, we
use the density relative to the mean density around each galaxy in a
given sample, 

\begin{equation}
\dpx=\rho_{\rm gal}/\langle{\rho}_{\rm gal}\rangle-1,
\end{equation}

\noindent where $\rho_{\rm gal}$ is the density in galaxies around
each object and $\langle{\rho}_{\rm gal}\rangle$ is the mean density
around all galaxies, as opposed the mean density of galaxies. Thus,
positive and negative values of $\dpx$ indicate galaxies that live in
higher or lower densities relative to the mean for that type of
galaxy.

The purpose of this paper is to quantify any correlations between the
properties of star forming central galaxies and their large-scale
environments. The scatter of star formation rates for galaxies on the
SFMS is high, and thus weak correlations with $\rhob$ can be easily
obscured by noise and limited statistics. To boost the signal-to-noise
of the measurement, we measure the mean environment as a function of
galaxy property, rather than the traditional method of binning
galaxies by $\rhob$ and calculating the mean of the galaxy property
within that bin. This method was used by \cite{hogg_etal:03} to
quantify the relationship between environment and galaxy luminosities
and colors.

\subsection{Numerical Simulations and Theoretical Models}
\label{s.sims}

As in Papers I and II, we will compare the results from the group
catalog to expectations from dark matter halos. Here we utilize two
simulations.  Most of our theoretical predictions use the `Chinchilla'
simulation, also used in the previous installments. The box size is
400 \hmpc\ per side, evolving a density field resolved with $2048^3$
particles, yielding a mass resolution of $5.91\times 10^8$ \hmsol. The
cosmology of the simulation is flat \lcdm, with $\om=0.286$,
$\sigma_8=0.82$, $h=0.7$, and $n_s=0.96$.  Halos are found in the
simulation using the Rockstar code of \cite{rockstar} and Consistent
Trees (\citealt{consistent_trees}) is used to track halo growth. Halo
masses are defined as spherical overdensity masses according to their
virial overdensity. The second simulation is smaller but with higher
mass resolution. This simulation was performed using the TPM code of
\cite{white:02}, and first presented in \cite{wetzel_white:10}. This
simulation has a box size of 250 \hmpc\ per side with 2048$^3$
particles, yielding a mass resolution 4 times higher than
Chinchilla. Halo finding uses the friends-of-friends algorithm with a
linking length of 0.18. This simulation will be used to track merger
histories of mock galaxies, as we will discuss in the next subsection.

The compare simulation results to galaxy results binned as a function
of environment, we measure the density around each halo in the
simulation in the same manner as for the galaxies. Using the
halo occupation distribution (HOD) fitting results of
\cite{zehavi_etal:11} from the SDSS Main galaxy sample, we populate
the simulation with galaxies that match the density and clustering of
each of our volume-limited samples. Using the distant-observer
approximation and the $z$-axis of the box as the line-of-sight, the
top-hat redshift-space galaxy densities are measured around each halo. 

We use the relation between central-$\mgal$ and $\mhalo$ shown in
Paper I (Figure 3 in that paper) to select halos to compare to
measurements at fixed central $\mgal$. This manner of selecting halos
does not include any scatter in the stellar mass-to-halo mass
relation, but in tests we have found that including scatter does not
change our results\footnote{To perform this test, we assign galaxies
  to dark matter halos using the mean SHMR, then shift the galaxy
  masses randomly using a Gaussian deviate with a width of 0.18 dex in
  $\log\mgal$. Then halos are selected by their {\it stellar mass}
  rather than $\mhalo$, to compare to observations. The correlations
  explored in these models---for example, the correlation between
  $\mdot$ and $\rhop$---vary weakly with halo mass. Thus, the
  introduction of scatter between halo mass and stellar mass does not
  impact the results.}. To test the hypothesis that halo assembly bias
is imprinted onto the properties of the galaxies, it is necessary to
make theoretical models that map halo properties onto galaxies
properties at fixed $\mhalo$. For comparisons to our three observable
galaxy properties---SFR, $\rexp$, and $\ns$---we map these properties
onto three different properties of dark matter halos.

\begin{figure*}
\psfig{file=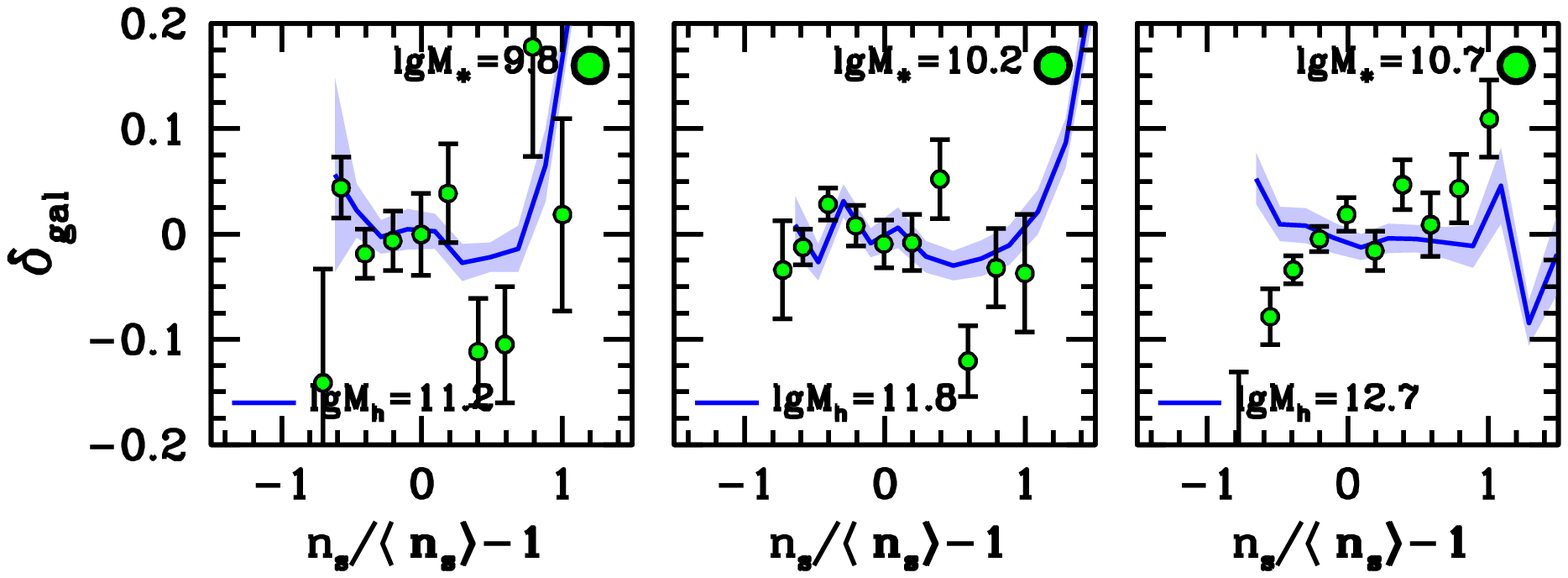,width=1\linewidth}
\vspace{-11cm}
\caption{ \label{mdot_ns_inv} Comparison between measurements and
  models for the correlation between $\ns$ and $\rhop$ for central
  galaxies. The points with errorbars are the same as those presented
  in Figure \ref{mdot_inv_all}. The curves are the results from halo
  abundance matching models. First, central galaxy stellar mass is
  mapped onto halo mass using the empirical relation from the group
  catalog. In each bin in halo mass, halo merger activity is mapped
  onto $\ns$ using the same procedure in Figure \ref{mdot_sfr_inv} for
  SFR (see text for details on how `merger activity' is defined and
  calculated). We assume a lognormal distribution of $\ns$ values
  with a dispersion that increases from 0.13 dex to 0.18 dex across
  the three galaxy mass bins. The solid curves
  represent the results from dark matter halos assuming no scatter in
  the relation between $\fmerge$ and $\rexp$. Because there is no
  clear correlation in the theoretical model, we do not include
  additional models that incorporate scatter in this relation. }
\end{figure*}

\begin{itemize}[leftmargin=0.1in]
\item Halo growth from $z=0.1$ to $z=0$, which we will refer to as
  $\mdot$. We use this halo property to assign SFR values to halos at
  fixed halo mass.
\item Halo angular momentum, parameterized through the dimensionless
  spin parameter $\lambda$, as defined by
  \cite{bullock_etal:01_spin}. We use this halo property to assign
  values of $\rexp$ to halos at fixed mass.
\item Galaxy merger history. We use the fractional amount of stellar
  mass in a mock galaxy accreted from galaxy mergers to map values of
  $\ns$ onto mock galaxies. We define this quantity as $\fmerge$.
\end{itemize}

To a reasonable approximation, the baryonic accretion rate onto a dark
matter halo is simply $f_b\times \mdot$, where $f_b$ is the universal
baryon fraction and $\mdot$ is the dark matter accretion rate onto the
halo (\citealt{behroozi_etal:13, moster_etal:13}). For low mass halos,
which shock heating is not efficient, gas will be accreted `cold' sink
to the center of the halo in a dynamical time (\citealt{keres_etal:05,
  keres_etal:09, dekel_birnboim:06}). Once at the halo center, the gas
should accrete onto the central galaxy and supplement the gas
reservoir from which stas are created. For higher mass halos, where
gas is no longer accreted cold, the situation is more complex but the
overall baryonic accretion rate will still follow the dark matter
accretion rate.

Using this relation between baryonic accretion rate and $\mdot$, we
can use the abundance matching ansatz to make theoretical models in
wich central galaxy sSFR is correlated with the growth of the dark
matter halo. In the simplest of such models, we assume no scatter
between sSFR and $\mdot$. In such a model, at fixed halo mass (and thus
$\mgal$), the halo with the highest growth rate contains the galaxy
with the highest sSFR, and on down the rank-ordered list. This is
analogous to the age-matching model of \cite{hearin_watson:13}, only for
sSFR rather than galaxy color. Although some fraction of galaxies at
any $\mgal$ are quiescent, in our model, all halos are available to
contain star-forming central galaxies. This means that star-forming
halos are not a `special subset' of host halos. Halos with quiescent
central galaxies represent a random subset of host halos. This is
backed up by the results of Papers I and II, in which the fraction of
quiescent central galaxies in independent of environment. 

Incorporating scatter in the $\mdot$-SFR relation is relatively
straightforward given that we assume a lognormal distribution in
SFR. See Appendix \ref{s.app_A}.

The theoretical models for assigning values of $\rexp$ to halos
follows in analogous fashion. We assume a lognormal distribution of
$\rexp$ values with a scatter of 0.2 dex independent of $\mgal$. In a
given bin of $\log\mhalo$, halos are ranked according to their angular
momentum, expressed through the dimensionless spin parameter $\lambda$
(\citealt{bullock_etal:01_spin}), which expresses the ratio between
the halo angular momentum and the angular momentum if the matter was
all in circular orbits. The quantity is calculated during the halo
finding process by the Rockstar algorithm.

Our procedure for constructing abundance-matching models for $\ns$
follows the same outline. We assume a lognormal distribution of
$\ns$. The scatter in $\ns$ increases from low to high $\mgal$,
however. Over the three bins in $\log\mgal$, the scatter in $\log\ns$
is 0.13, 0.155, and 0.18 dex. To calculate $\fmerge$, we first
identify all $z=0$ distinct halos (i.e., halos that house central
galaxies). With this list, we follow the evolution of these halos
forward in time starting at $z=1$, when the typical mass of a Milky
Way sized galaxy is only $\sim 20\%$ of its present-day value. Stellar
masses are assigned to each halo at each redshift independently, using
abundance matching and the stellar mass functions measured by at each
redshift range (see \citealt{hahn_etal:17_tq} for details of this
model).

As smaller halos are accreted onto larger halos, mergers take place
when a satellite is no longer identifiable as its own
halo. \cite{wetzel_white:10} determined that satellite disruption
occurs when a subhalo is stripped of 97 to 99\% of its mass. This
criterion, combined with abundance matching, gives results consistent
with observations of spatial clustering and the fraction of galaxies
that are satellites. For low-mass galaxies that live in $10^{11.2}$
$\msol$ halos, our procedure may overestimate the {\it number} of
minor mergers that occur because we are unable to track all halos
below $10^{11}$ $\msol$ down to 1\% of their mass at the time of
accretion, but given the slope of the stellar-to-halo mass relation,
the overal contribution of these galaxies to the $z=0$ stellar mass of
a galaxy is likely to be small.

As the evolution of each halo is followed, the total stellar mass of
satellite galaxies that have merged with the parent galaxy is summed
up. We define $\fmerge$ as the ratio between this mass and the $z=0$
stellar mass within the halo, as defined by abundance matching once
again. 

\begin{figure*}
\psfig{file=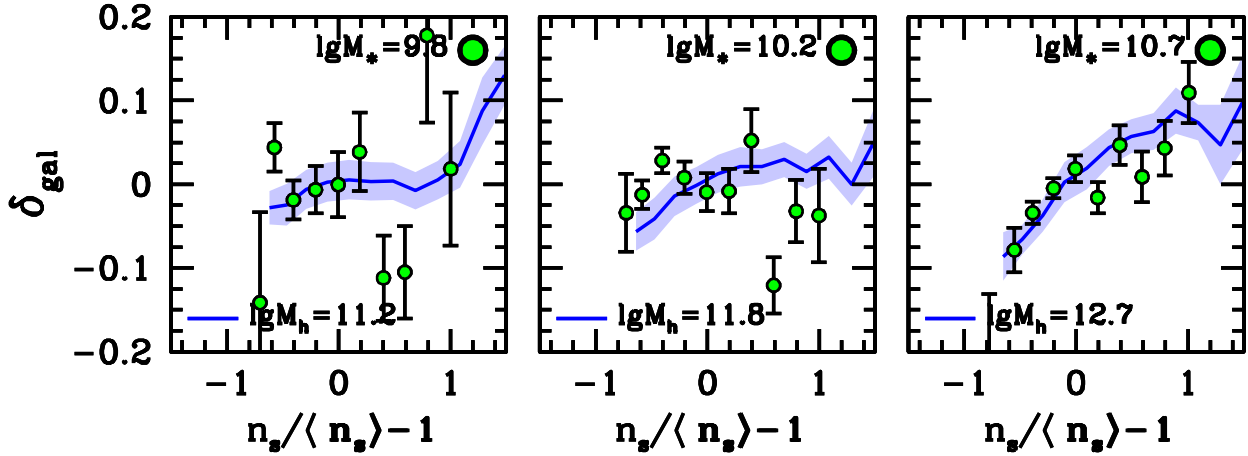,width=1\linewidth}
\vspace{-11cm}
\caption{ \label{mdot_ns_angmom} Same as Figure \ref{mdot_ns_inv}, but
  now the curves show a theoretical model in which halo spin parameter
  $\lambda$ is mapped onto $\ns$. There is no scatter between the two
  in this comparison. }
\end{figure*}

\section{Results}

\subsection{Do the Properties of Star-Forming Central Galaxies
  Depend on Large-Scale Environment?}

According to Figure \ref{mdot_inv_all}, the answer depends on the
property in question. As detailed in \S 2, all results in this section
are restricted to pure central galaxies that lie on the SFMS.  Each
row shows results for a different galaxy property. The bottom row
shows results when binning galaxies by $\rexp$, the scale radius of
the exponential fit to each galaxy's light profile. The middle row
shows results for $\ns$, the best-fit Sersic index to the galaxy light
profile. The top row shows results when binning galaxies by sSFR. The
columns show bins in stellar mass. From left to right, the bins are
$\log\mgal=[9.7,10.0]$, $[10.1,10.5]$, $[10.7, 10.9]$. Wide bins are
necessary to increase the statistical power of the samples. In each
panel, the $x$-axis is the galaxy property relative to the mean. Due
to the width of the bin, the mean of a galaxy property can evolve
significantly from the low-mass end of the bin to the high-mass
end. To prevent biases from this evolution, the mean galaxy property
is first calculated in 0.1-dex sub-bins of $\log\mgal$, and the each
galaxy's properties are with respect to the mean in the
sub-bin. Errors represent statistical errors in the mean.

The top row shows results when binning galaxies by sSFR. In each
panel, there is a clear correlation between
$\ssfr/\langle\ssfr\rangle$ and $\rhop$ such that galaxies that are
stronger than average star-formers live in lower densities, while
below-average star-formers live in higher densities. The slope of this
correlation is monotonically shallower with higher $\mgal$, but in
each panel there is a statically significant correlation: using a
$\chi^2$ statistic to test the consistency of each panel's result with
a straight line yields $\chi^2$ values of 23.6, 31.6, and 22.3 for
each panel from left to right, respectively, for 10 data points in
each panel. The error bars are smaller for higher $\mgal$ bins due to
the increased volume available for higher-mass samples.  We will
discuss this correlation in the context of dark matter halo growth in
the following subsection.

The middle row shows results when binning galaxies by $\ns$. There is
no clear correlation between $\ns$ and $\rhop$ for$\mgal\la 10^{10.5}$ $\msol$,
but at high masses a statistically significant
correlation exists. In the far-right panel, galaxies with higher $\ns$---i.e.,
galaxies with more elliptical and less disky morphology---live in slightly
higher than average density environments. Galaxies that are more
disk-dominated, however, live in significantly lower densities than
average. The $\chi^2$ test described above yields values of 16.8,
21.0, and 37.9 for the panels from left to right, respectively. The
$\chi^2$ of 21.0 for the middle panel is mostly driven by the datum at
$\ns/\langle\ns \rangle=0.6$, without which the $\chi^2$ is 8.0. The
large $\chi^2$ for the high mass bin is distributed more evenly
in the data. Removing the far left datum, which has $\rhop=-0.22$,
reduces the $\chi^2$ to 32. 

The bottom panel shows results for $\rexp$. In all panels, there is no
clear dependence of $\rhop$ on $\rexp/\langle\rexp\rangle$. The
$\chi^2$ test yields values, from left to right, of 6.8, 11.7, and
10.7. There is a slight positive slope in the high-mass bin. A line
with a slope of 0.05 yields a $\Delta \chi^2=4.3$ with respect to a
straight line, but a straight line fit is statistically reasonable
given 10 data points. We will discuss these results in the context of
halo angular momentum in subsequent subsections.

\begin{figure*}
\psfig{file=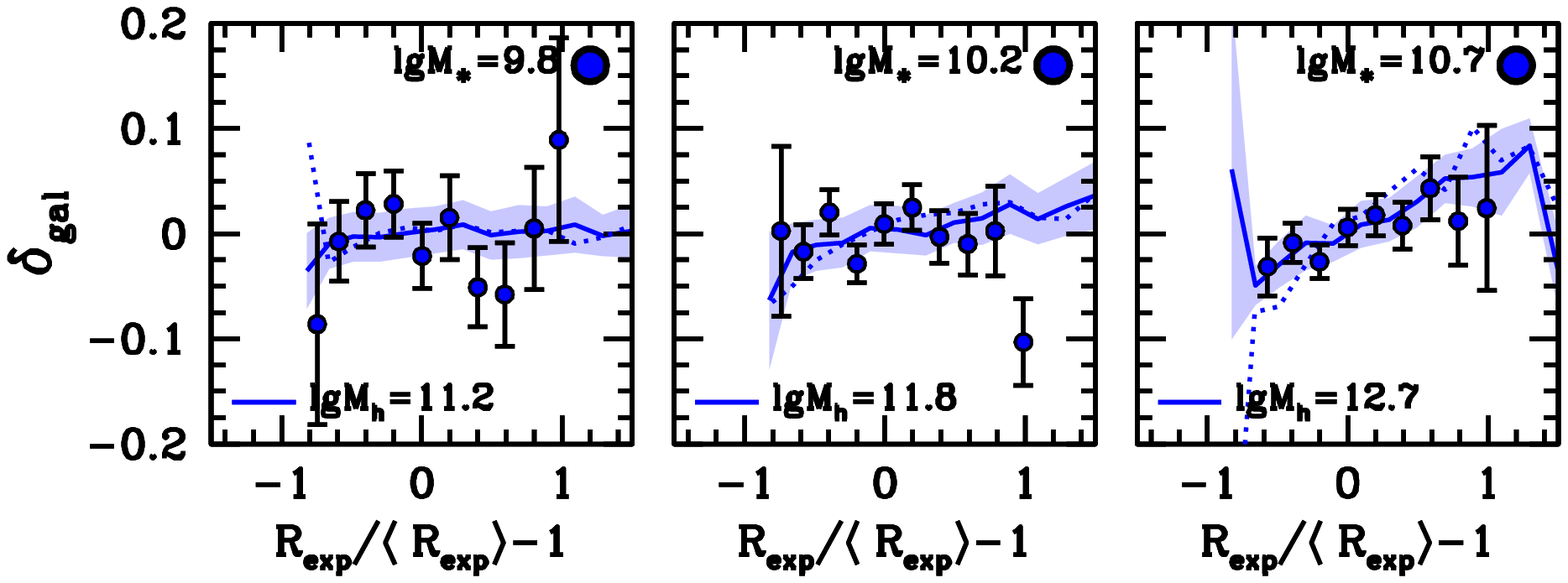,width=1\linewidth}
\vspace{-11cm}
\caption{ \label{mdot_rexp_inv} Comparison between measurements and
  models for the correlation between $\rexp$ and $\rhop$ for central
  galaxies. The points with errorbars are the same as those presented
  in Figure \ref{mdot_inv_all}. The curves are the results from halo
  abundance matching models. First, central galaxy stellar mass is
  mapped onto halo mass using the empirical relation from the group
  catalog. In each bin in halo mass, halo angular momentum is mapped
  onto $\rexp$ using the same procedure in Figure \ref{mdot_sfr_inv}
  for SFR. We assume a lognormal distribution of $\rexp$ values with a
  dispersion in $\log\rexp$ of 0.2 dex. The dotted curves represent
  the results from dark matter halos assuming no scatter in the
  relation between $\lambda$ and $\rexp$. The solid curve assumes a
  scatter in $\log\rexp$ at fixed $\lambda$ of 0.17 dex. The assembly
  bias of halo angular momentum does not become significant in this
  relation until $\mhalo\ga 10^{12}$ $\msol$. The model with no scatter
  produces too steep a slope to be consistent with the data, but the
  model with 0.17 dex of scatter is a better fit than no relation at
  all, yielding a $\Delta\chi^2$ of 3.2 with respect to a straight line. }
\end{figure*}

\subsection{Does Halo Growth Rate Correlate with Galaxy Growth Rate?}

Using the method described in \S 2.3, we match halo growth rate to
galaxy sSFR. In this scenario, the fastest growing halos have the highest
star formation rates, while the slowest growing halos (or
negatively-growing halos) have the lowest sSFR values. Because there
is a correlation between halo growth rate and halo environment, this
will impart a strong correlation between $sSFR$ and $\rhop$ in the. It
is important to note the results of Papers I and II, which imply that
quenching is a stochastic process with respect to halo growth rate,
especially for halos with $\mhalo\la 10^{12.5}$. Thus, star-forming
halos are likely not a `special subset' of dark matter halos, and we
can draw from the full population of halos to make predictions.

Figure \ref{mdot_sfr_inv} compares two theoretical models to the data
presented in the top row of Figure \ref{mdot_inv_all}. The dotted
curves show one model in which $\mdot$ is mapped onto $\log$~sSFR
assuming no scatter between the two quantities. The solid curves are
the results of a model in which the scatter between these two
quantities is 0.25 dex. Both models yield results that are
qualitatively in good agreement with the data: the models predict and
inverse correlation between star formation rate and $\rhop$, with a
slope that monotonically decreases with increases $\mgal$. The slope
of the correlation predicted in the no-scatter model is too steep
relative to the data, especially for the slowest-growing halos. The
model that incorporates scatter, however, is in excellent agreement
with the data. The value of 0.25 dex was obtained by finding the
scatter that yielded the lowest $\chi^2$ when comparing the model to
the data, with a value of 28 for 30 data points. This model yields a
$\Delta\chi^2$ of 21 with respect to a model with no correlation (but
has the same errors as the simulation). A scatter of 0.25 dex in
$\log$~sSFR is approaching the overall scatter of in the SFMS of 0.28
dex, but even with this amount of scatter at fixed $\mdot$, the model
still creates a significant correlation between sSFR and
$\rhop$. These values imply a correlation coefficient of $r=0.63$
between $\mdot$ and $\log$~sSFR.

\subsection{Does merger activity correlate with galaxy light profile?}

Figure \ref{mdot_ns_inv} shows the $\ns$-$\rhop$ relation first
presented in Figure \ref{mdot_inv_all}, but now with a comparison to
the model in which $\fmerge$ is abundance matched onto $\ns$ at fixed
$\mhalo$. Error bars are calculated by jackknife sampling of the
simulation volume into 8 sub-volumes. The slightly larger error bars
in this comparison, relative to those seen in Figure
\ref{mdot_sfr_inv} and what we will see in Figure \ref{mdot_rexp_inv}
are due to the smaller box size of this simulation. 

In all three panels, the model shows no evidence of a correlation
between $\fmerge$ and $\rhop$, and thus yields no correlation between
$\ns$ and $\rhop$. In our fiducial model, we incorporate 0.2 dex of
scatter in $\log\mgal$ at fixed $\log\mhalo$. This value is consistent
with recent measurements (e.g., \citealt{reddick_etal:13, zu_mandelbaum:15,
  tinker_etal:17_boss}). Physically, the amount of merging a galaxy has
over its lifetime may contribute to this scatter, but this is not
reflected in our fiducial implementation. Thus, we have run an
additional model in which there is no scatter at $z=0$. The results
are unchanged, verifying that our fiducial model is not affected by
uncorrelated scatter. 

The parameter $\ns$ need not be correlated only to $\fmerge$, but it
is difficult to find another halo parameter that could match the
signal measured in Figure \ref{mdot_ns_inv}. This is due to the fact
that assembly bias created by most halo parameters is maximal at low
halo masses---$\mhalo\la 10^{11}$ $\msol$. These parameters include
$\cvir$, $\zhalf$, or short-term halo growth, $\mdot$. However, halo
spin parameter $\lambda$ yields a rather different assembly bias
signal than these other parameters (see, e.g,
\citealt{gao_white:07}). For $\lambda$, the assembly bias signal
actually gets larger with higher halo mass, and goes away completely
at $\mhalo\la 10^{12}$ $\msol$. Figure \ref{mdot_ns_angmom} shows the
comparison between measurements and an abundance matching model which
maps $\lambda$ onto $\ns$, with no scatter, at fixed $\mhalo$. This
comparison is quite favorable, matching the slope of the observed
correlation at $\mgal\sim 10^{10.8}$ and showing little-to-no
correlation in the lower mass bins.  For comparison, in the right-hand
panel, we also show a model with 0.16 dex of scatter between $\log\ns$
and $\log\lambda$. The slope of the correlation is notably shallower,
although it is difficult to distinguish between them given the noise
in the data. Formally, when combining the jackknife errors from the
simulation to the errors in the SDSS centrals, the minimum $\chi^2$ is
achieved with a scatter of 0.11 dex, with a $\Delta\chi^2<1$ range
of $0.08\le\sigma(\log\ns|\log\lambda)\le 0.14$. These $\chi^2$ values
add the errors in the model and the errors in the data in quadrature.

\subsection{Does Halo Angular Momentum Correlate with Galaxy Disk Size?}

As already shown in the previous subsection, the assembly bias signal
created by halo spin has minimal amplitude at $\mhalo\la 10^{12}$
$\msol$, but becomes measurable for galaxies that live in higher mass
halos. For $\ns$, there is a clear correlation with environment at
$\mgal\sim 10^{10.8}$ $\msol$, and this correlation is consistent with
a model in which $\ns$ is strongly correlated with $\lambda$. For
$\rexp$, the observational situation is less clear. In our high-mass
galaxy bin, there is a measurable slope in the $\rexp$-$\rhop$ data,
but the statistical significance is low, given that a straight-lin fit
yields a $\chi^2$ of 10.7 for 10 data points.

Figure \ref{mdot_rexp_inv} compares these data to an abundance
matching model that maps $\lambda$ onto $\rexp$ at fixed $\mhalo$. As
described above, this model assumes a lognormal distribution of
$\rexp$ with a scatter of 0.2 dex. As expected from Figure
\ref{mdot_ns_angmom}, the assembly bias signal of this model is
minimal in the first two stellar mass bins. But in the highest $\mgal$
bin, the model yields a measurable signal. The dotted curves in Figure
\ref{mdot_rexp_inv} show the model with no scatter between $\lambda$
and $\rexp$. The correlated between $\rexp$ and $\rhop$ created in
this model is stronger than that seen in the data. The solid curves
show a model with 0.17 dex in scatter in $\log\rexp$ at fixed
$\lambda$. This model yields a slightly better fit to the data than a
model with no correlation. To put the comparison on equal footing, we
use the errors in the model to calculate a new $\chi^2$ for the
no-correlation model. This reduces the $\chi^2$ from 10.7 to 3.9 for
this model. The model with 0.17 of scatter in the relation yields a
$\chi^2$ of 1.4.

\section{Discussion}

Papers I and II of this series demonstrated that large-scale
environment---and, by extension, halo growth history---plays a limited
impact on whether a central galaxy is quenched. In this paper we have
restricted our analysis to central galaxies that lie on the
star-forming sequence, allowing us to examine properties that are
unique to such galaxies; the star-formation rates, disk sizes, and
light profiles. As with Paper I and II, previous investigations have
focused on how assembly bias might impact either galaxy bimodality
(see, e.g., \citealt{lacerna_etal:14, lin_etal:16}) or the full galaxy
population (e.g., \citealt{tinker_etal:08_voids,
  zentner_etal:16}). This is the first study to look at secondary
properties within the set of star-forming galaxies.

\subsection{Assembly bias and star formation rates}

Our results indicate that, at fixed stellar mass, central galaxies on
the SFMS have higher star formation rates in lower density
environments. These data are consistent with a model in which sSFR is
correlated with near-term halo growth rate. This is a detection of
assembly bias within this class of galaxies. It demonstrates
consistency between the assumptions of the abundance matching
model---namely, that galaxy growth should be correlated with halo
growth---and the properties of observed star-forming galaxies. 

Extrapolated to high redshift, this result implies that the total
stellar mass of the galaxy is related to the formation history of its host
halo. Using redshift-dependent abundance matching,
\cite{behroozi_etal:13} calculated the efficiency of converting
accreted baryons into stars as a function of both time and halo
mass. For halos less massive than $10^{12}$ $\msol$ at $z=0$, this
efficiency is lower at higher redshift than it is today. Thus, halos
that form early accrete most of their baryons when this conversion
efficiency is low, and will form galaxies that are less massive than
late-forming halos of the same mass that accrete most of their mass at
when efficiencies are higher. \cite{tinker:17} demonstrated that this
is a source of scatter in the stellar-to-halo mass relation.

This is, however, the opposite of the assembly bias described in
\cite{lim_etal:16} and measured in the GAMA survey by
\cite{tojeiro_etal:16}. In \cite{lim_etal:16}, the ratio of
$\mgal/\mhalo\equiv f_c$ is a proxy for halo formation time, with
halos with higher $f_c$ forming earlier. This effect is also seen in
hydrodynamic simulations (\citealt{matthee_etal:17}). In these models,
halos that form early accrete significant amounts of gas early, and
this gas therefor has a longer timescale over which to form
stars. These two scenarios make mutually exclusive predictions for the
halos around these galaxies. At fixed halo mass, the abundance
matching model predicts that later-forming halos have larger galaxies,
while in the SAMs and hydro simulations, early-forming halos have
larger galaxies. Thus, at fixed central galaxy stellar mass, abundance
matching predicts that halo mass will increase as you go from
sSFR/$\langle$sSFR$\rangle sim -1$ to $+1$. The other models will
predict the opposite trend. Galaxy-galaxy lensing or satellite
kinematics may be able to provide discriminating information.

\subsection{The role of spin in galaxy formation}

The correlation between other SFMS galaxy parameters---$\ns$ and
$\rexp$---is less clear. At galaxy masses $\mgal\la 10^{10.4}$
$\msol$, the data show no correlation between large-scale environment
and these properties. However, for our high-mass galaxy bin,
$\log\mgal=[10.6, 10.9]$, the data show a positive correlation between
both parameters and $\rhop$. For $\ns$, the correlation is highly
significant, while for $\rexp$, a model with no correlation is still a
satisfactory description of the data. These trends in the data are
consistent with a model in which both parameters are positively
correlated correlated with halo angular momentum, parameterized by the
the spin, $\lambda$. Halo properties that are tightly linked with halo
growth history, $\zhalf$, $\cvir$, and $\mdot$, show strong
assembly bias at low halo masses, $\mhalo\la 10^{12}$ $\msol$ at
$z=0$. This is consistent with how the observed correlation between
SFR and $\rhop$ changes with $\mgal$. However, halo spin shows a
distinctly different relationship with large-scale environment at
fixed mass. \cite{bett_etal:07} show that spin is {\it uncorrelated}
with large-scale environment at $\mhalo\la 10^{12}$ $\msol$, but at
higher halo masses, halos with higher spin exhibit stronger
clustering. The only other halo property that shows this trend of
stronger assembly bias at higher halo masses is the total number of
subhalos within the parent halo (\citealt{croft_etal:12, mao_etal:17}). 

The similarity between the assembly bias signals of $\lambda$ and
amount of substructure suggest that the two are correlated. In
traditional tidal torque theory, angular momentum is imparted on a
halo at early times, when structure is still linear, and is related to
the distribution of matter in the initial density field (see, .e.g,
\citealt{porciani_etal:02}). Alternatively, \cite{vitvitska_etal:02}
proposed a model for the origin of halo spin through accretion of
substructure. Major mergers spin up halos significantly, while angular
momentum is also accumulated through lower-mass halo mergers.

However, if merging and spin are correlated, and through this
correlation yield the same assembly bias signal, why does our merger
model fail to produce a correlation between $\fmerge$ and $\rhop$?
Perhaps the smaller simulation volume, $250^3$ $($Mpc/$h)^3$, limits
our ability to make a clear detection. Or it is possible that our
chosen statistic, $\fmerge$, is not optimal to detect the assembly
bias signal. Galaxy mergers are quite different than halo accretion
events, and given the steepness of the stellar-to-halo mass relation
at low masses, it is possible for a halo to have a number of accretion
events without building up much stellar mass through such events (see,
e.g., \citealt{maller:08}). Or, as pointed out by \cite{mao_etal:17},
even though the two properties are correlated, it is still possible
that the scatter in this correlation eliminates any assembly bias
signal in one of the two properties.

\cite{rodriguez_gomez_etal:17} investigated the correlation between
galaxy morphology, spin, and merger activity in the Illustris
simulation. In their results, spin does correlate with morphology, but
only for lower-mass galaxies. Spin may influence galaxy properties at
low mass in our SDSS samples as well, but since $\lambda$ does not
yield and assembly bias in the host halos of these galaxies, there is
no observational signature of such a correlation. The statistic probed
in \cite{rodriguez_gomez_etal:17} was the fraction of total kinetic
energy in the galaxy contributed by rotational motion, which is not
possible to measure directly in the full group catalog. 

We also note that the results of Paper I appear similar to those of
$\ns$ and $\rexp$; at low galaxy masses, there is no correlation
between the quenched fraction of central galaxies and
$\rhop$. However, at higher stellar masses, there is indeed a small
but non-zero slope in the correlation such that central galaxies in
higher density regions are more often quenched. This is consistent
with the assembly bias yielded from a correlation with halo spin, but
the implication would be that {\it higher} spin halos are more likely
to be quenched. At first glance, this result would seem to challenge
the traditional orthodoxy of galaxy formation within dark matter
halos---namely, that high-spin halos would form rotationally supported
galaxies. As noted by \cite{vitvitska_etal:02}, however, if spin is
indeed created by mergers, the merger activity may cause galaxy
transformation. The merger scenario for galaxy quenching has come into
question, as hydrodynamical simulations suggest that, without the
presence of an active post-merger feedback mechanism, star-formation
is likely to be restarted after the merger is complete
(\citealt{pontzen_etal:17}). But mergers may still temporarily quench
galaxies, or the induced quenching may be permanent for some small
fraction of merger events.

Of course, we cannot rule out the possibility that the agreement
between the halo spin abundance matching models and the data is simply
coincidence. But the results here indicate that it further
investigation of the secondary properties of passive galaxies---their
velocity dispersions, size, and light profiles---may elucidate the
processes that caused their transformation to the red sequence.

\subsection{Emission-Line Galaxy samples as cosmological probes}

The detection of assembly bias in star forming objects may have
implications for the use of such objects as tracers of the dark matter
density field. The emission line galaxy (ELG) has been situated as the
cosmological workhorse for the next generation of galaxy redshift
surveys. Data are already being taken on a cosmological sample in the
eBOSS program (\citealt{dawson_etal:15}; see
\citealt{raichoor_etal:17} for details of the ELG selection). Assembly
bias in ELG samples would alter both their large-scale bias and the
shape of their clustering, relative to a model that assumes that halo
mass is the only property that determines their occupation
(\citealt{sunayama_etal:16}). This is unlikely to bias measurements of
baryon acoustic oscillations, but may have an impact on efforts to
use clustering as a probe of the growth rate, neutrino masses, and
non-Guassianity. Given the high precision expected from the clustering
measurements of ELG samples, further investigation of the possible
impact of the type of assembly bias measured here is warranted.


\section*{Acknowledgements}
\noindent The authors wish to thank Michael Blanton, Rita Tojeiro, and
Risa Wechsler for many useful discussions. The authors thank Matthew
Becker for providing the Chinchilla simulation used in this work. The
Chinchilla simulation and related analysis were performed using
computational resources at SLAC. We thank the SLAC computational team
for their consistent support. JLT acknowledges support from National
Science Foundation grant AST-1615997. AW was supported by a
Caltech-Carnegie Fellowship, in part through the Moore Center for
Theoretical Cosmology and Physics at Caltech, and by NASA through
grant HST-GO-14734 from STScI.

\appendix
\begin{figure*}
\psfig{file=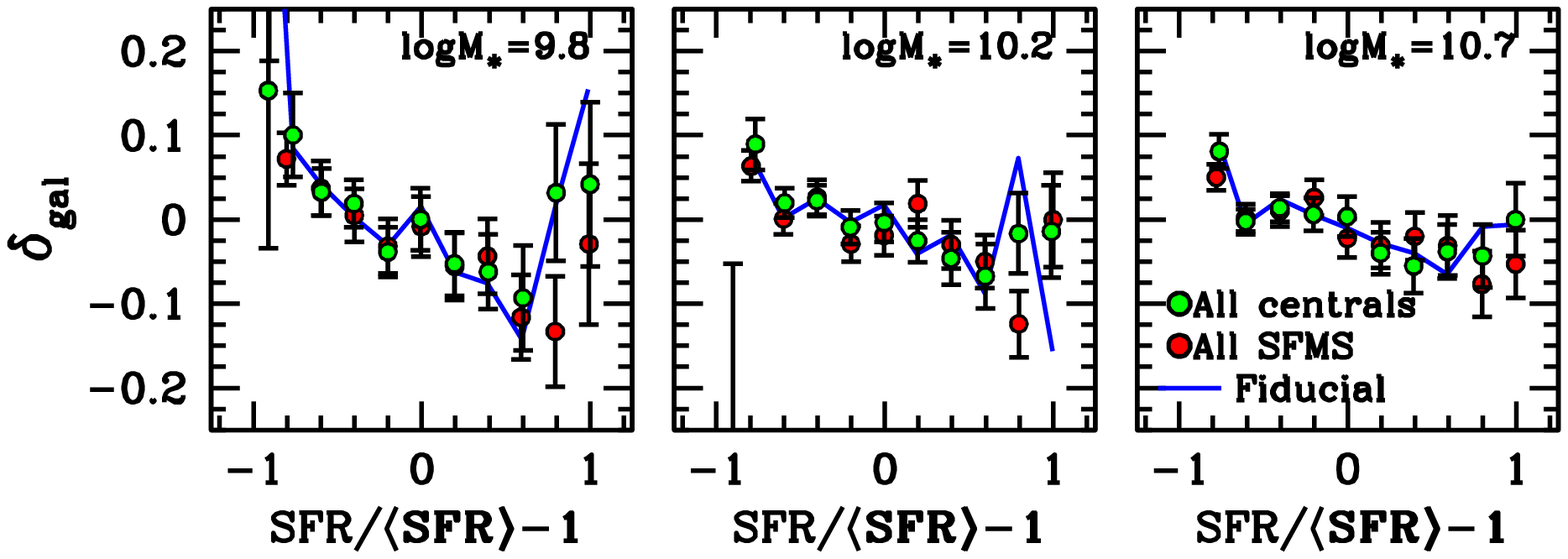,width=1\linewidth}
\vspace{-11cm}
\caption{ \label{mdot_inv_tests} Tests of the measurements of the
  correlation between sSFR and $\rhop$. Our fiducial make two cuts on
  the full sample of star-forming central galaxies. First, our results de-weight
galaxies in the transition region of the sSFR distribution, where
galaxies may be in the process of quenching their star
formation. Second, we only use pure central galaxies, i.e., central
galaxies with $\psat<0.01$. Our fiducial results are shown with solid
blue curves. Results using the full sample of central galaxies are
shown with the green circles. Results using all star-forming galaxies,
without any de-weighting, are shown with the red circles. Results are
un affected by these cuts on the sample. }
\end{figure*}

\begin{figure*}
\psfig{file=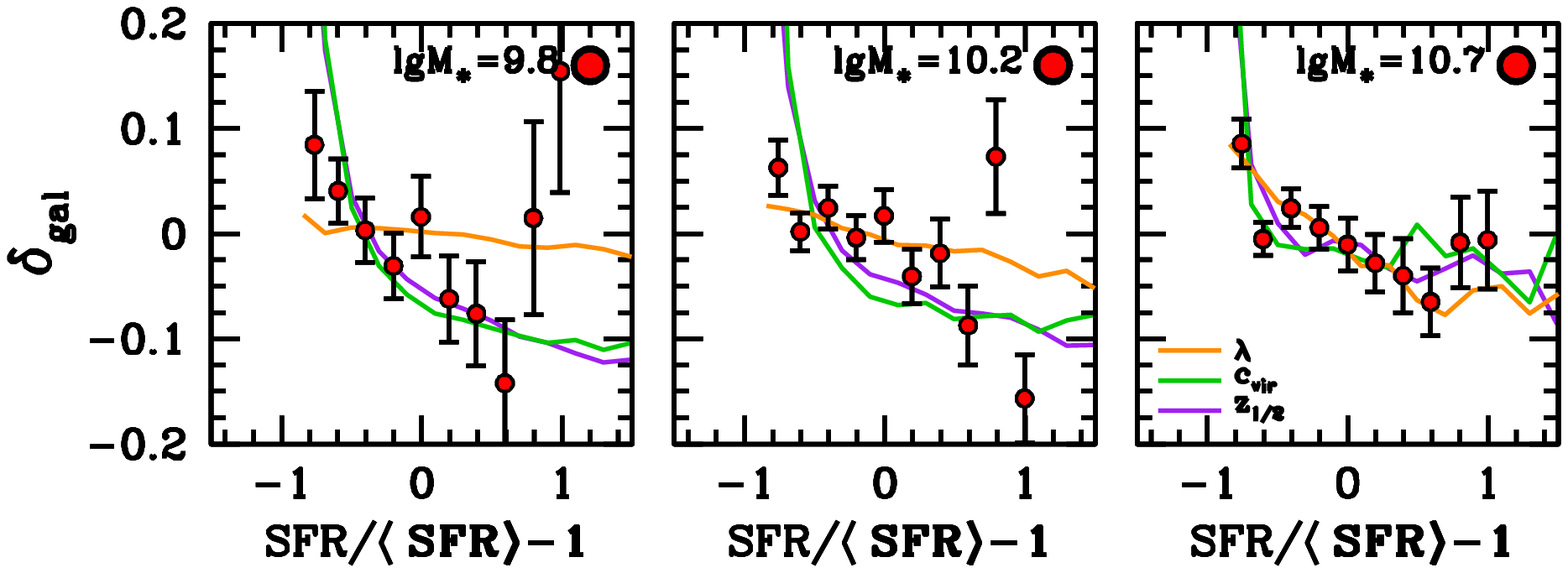,width=1\linewidth}
\vspace{-11cm}
\caption{ \label{mdot_sfr_mix} The mean large-scale density around
  central galaxies as a function of their relative star formation rate
  in several bins in stellar mass. For each bin in $\log\mgal$, low
  star-formers reside in higher densities, which above-average
  star-formers live in lower densities. However, the correlation gets
  progressively weaker with higher $\mgal$. In each panel, we compare
  to models in which halo growth rate is matched to SFR such that the
  fastest growing (or youngest) halos have the highest star formation
  rates. The first two curves rank-order the halos by growth over a
  redshift interval indicated in the key. The orange curve shows the
  results when rank-ordering halos by $\zhalf$. }
\end{figure*}

\section{Abundance Matching for Star Formation Rate}
\label{s.app_A}

First, we assume the SFMS is a lognormal. Thus, the cumulative rank
$i_{\rm rank}$ at
any location in the distribution can be expressed as

\begin{equation}
\label{e.erf}
i_{\rm rank} = \frac{1}{2}\left[1 - {\rm erf}(x/\sqrt{2})\right]
\end{equation}

\noindent where $i_{\rm rank}$ is a normalized rank in the range $[0,1]$
and $x$ is defined as

\begin{equation}
\label{e.x}
x = \frac{SFR-\langle SFR\rangle}{\sigma_{\log SFR}}
\end{equation}

\noindent and we assume $\sigma_{\log SFR}$ is 0.3, independent of
stellar mass. For the case of no intrinsic scatter between $\mdot$ and
SFR, the rank-ordered list of halos can be matched to SFR by inverting
equation (\ref{e.erf}). To include intrinsic scatter, $\sigma_{\rm
  int}$, we assume that the {\it total} scatter in SFMS is 0.3,
and the value used in equation (\ref{e.x}) is 

\begin{equation}
\sigma_{\log SFR}^2 = 0.3^2 - \sigma_{\rm int}^2.
\end{equation}

\noindent Thus, after determining the SFR of each halo based on
abundance matching, each halo receives an additional $\log$ SFR drawn
from a Gaussian distribution with zero mean and $\sigma=\sigma_{\rm
  int}$.

\section{Testing Different Samples of Central Star-Forming Galaxies}
\label{s.app_B}

In Figure \ref{mdot_inv_tests}, we show measurements of the correlation
between SFR and $\rhop$ for different samples of star-forming
galaxies. Our fiducial sample contains only star-forming galaxies
likely to be on the SFMS. Thus we randomly remove galaxies with low
star formation rates in order to preserve the lognormal distribution
of SFR. Additionally, our fiducial sample contains only central
galaxies indicated as `pure' centrals by the group catalog. These are
centrals with $\psat>0.99$. Removing non-pure centrals only reduces
the overall sample size by $\sim 10\%$. 

In FIgure \ref{mdot_inv_tests} we compare our fiducial measurements to
those for samples in which we relax the restrictions on the
sample. The blue curve is the fiducial measurement from Figure
\ref{mdot_inv_all}. The green circles show measurements that include
all central galaxies, not just pure centrals. The red circles show
measurements for all star-forming galaxies down to a specific SFR of
$10^{-11}$ yr$^{-1}$. In both of these tests, the results are fully
consistent with the fiducial measurement.

\section{Comparison of SFR Results to Different Halo Properties}
\label{s.app_C}

Figure \ref{mdot_sfr_mix} compares the measurements of the SFR-$\rhop$
correlation to abundance matching models that use halo properties
other than $\mdot$ as a proxy for star formation. Here we replace
$\mdot$ with $\zhalf$, the redshift at which the halo reached half its
present-day mass, the concentration parameter $\cvir$, and the halo
spin parameter $\lambda$. For the first two halo properties, there is
a correlation between $\zhalf$, $\cvir$ and $\mdot$ (see, e.g.,
\citealt{wechsler_etal:02}). Thus, all of these halo properties yield
similar correlations. We note, however, that the {\it sign} of the
correlation is opposite to that of $\mdot$. In Figure
\ref{mdot_sfr_inv}, halos with the highest $\mdot$ had the highest
SFR. For $\cvir$ and $\zhalf$, halos with the lowest values of these
properties have the highest star formation rates. Note that there is
no scatter introduced in these comparisons. 

As expected from Figure \ref{mdot_ns_angmom} and \ref{mdot_rexp_inv},
the assembly bias signature created by a model that matches $\lambda$
to SFR does not compare favorably to the data. Here, again, we assert
an inverse relationship between spin and star formation rate, with no
scatter. 

For $\mdot$, We do find some dependence on the time baseline over
which $\mdot$ is calculated. The maximal assembly bias signal is found
for $\mdot$ calculated over a redshift baseline of $\Delta
z=0.8$. Smaller values of $\Delta z$ correspond to smaller assembly
bias signals. At baselines larger than 0.8, the assembly bias signal
is largely unchanged. We have not shown these results to preserve
clarity in the plot, but the results for $\Delta z=0.8$ are comparable
to those for $\cvir$.


\bibliography{../risa}

\label{lastpage}

\end{document}